\documentclass[a4paper]{jpconf}
\usepackage{graphicx,url}
\usepackage{amsmath,amssymb}
\usepackage{color}
\usepackage{subfig}
\usepackage{ulem}   

\begin{document}

\title{Performance of a Chirplet-based analysis for gravitational-waves from binary black-hole mergers}
\author{Satya Mohapatra$^{\dagger}$, Zachary Nemtzow$^{\dagger}$, \'Eric Chassande-Mottin$^{\ddagger}$ and Laura Cadonati$^{\dagger}$ } 
 \address{$^{\dagger}$  Physics Department, University of Massachusetts, Amherst MA 01003  (USA)\\
$^{\ddagger}$ APC, Univ Paris Diderot, CNRS/IN2P3, CEA/Irfu, Obs de Paris, Sorbonne Paris Cit\'e (France)}

\ead{satya@physics.umass.edu}

\begin{abstract}
The gravitational wave (GW) signature of a binary black hole (BBH) coalescence is characterized by rapid frequency evolution in the late inspiral and merger phases. For a system with total mass larger than 100 $M_{\odot}$, ground based GW detectors are sensitive to the merger phase, and the in-band whitened waveform is a short-duration transient lasting about  10-30 ms. For a symmetric mass system with total mass between 10 and 100 $M_{\odot}$, the detector is sensitive instead to the inspiral phase and the in-band signal has  a longer duration, between 30 ms - 3 s.  \textit{Omega} is a search algorithm for GW bursts that, with the assumption of locally stationary frequency evolution, uses sine-Gaussian wavelets as a template bank to decompose interferometer strain data. The local stationarity of sine-Gaussians induces a performance loss for the detection of lower mass BBH signatures, due to the mismatch between template and signal. We present the performance of a modified version of the Omega algorithm, \textit{Chirplet Omega}, which allows a linear variation of frequency, to target BBH coalescences. The use of Chirplet-like templates enhances the measured signal-to-noise ratio due to less mismatch between template and data, and increases the detectability of lower mass BBH coalescences. We present  the results of a performance study of \textit{Chirplet Omega} in colored Gaussian noise at initial LIGO sensitivity. \end{abstract}

\section{Motivations}

Gravitational wave burst searches~\cite{s5-all-sky-burst} are designed to search for short duration transients. Burst search algorithms typically look for excess signal energy in the time-frequency plane after filtering the GW strain data with a particular template bank. Current burst algorithms, such as {\it Omega}~\cite{chatterji05:_ligo} and {\it Coherent Waveburst}~\cite{cwb} use sine-Gaussians and a collection of Meyer wavelet bases respectively as templates. Matched filtering searches for the coalescence of compact binary systems  with total mass $2-25 M_{\odot}$~\cite{cbclowmass},  $25-100 M_{\odot}$~\cite{cbchighmass} and $> 100 M_{\odot}$~\cite{ringdown} have also been performed. These matched filtering algorithms use signal models such as the Post Newtonian approximation~\cite{owen99:_match}, the EOBNR analytical template family~\cite{eobnr} and quasinormal ringdown  modes,  respectively.

Inspiralling binary black holes are characterized by chirping GW signals, where the frequency changes rapidly in the later inspiral and the merger phases.  The duration of the signal in the most sensitive band of a ground based detector is dependent on the mass of the binary. For a system of total mass larger than 100 $M_{\odot}$, a ground based GW detector is sensitive to the merger phase, and the in-band whitened waveform is a short-duration transient lasting 10-30 ms. For a symmetric binary system with a total mass between 10 and 100 $M_{\odot}$, the detector is sensitive instead to the inspiral phase and the in-band signal has  a longer duration between 30 ms - 3 s. A chirplet based burst search has  been proposed in Ref.~\cite{chirplet-paper}, which combines the chirp characteristic of an inspiral with the existing Omega burst search. In this paper we explore further the use of the chirplet burst search and benchmark the detection performance for  BBH coalescences, compared to the standard Omega burst search at a constant false alarm rate.
 
\section{Chirplet  Definition}

\begin{figure}
\begin{tabular}{cc}
\subfloat[Part 1][A sample sine-Guassian.]{\includegraphics[width=.45\textwidth,height=0.25\textheight]{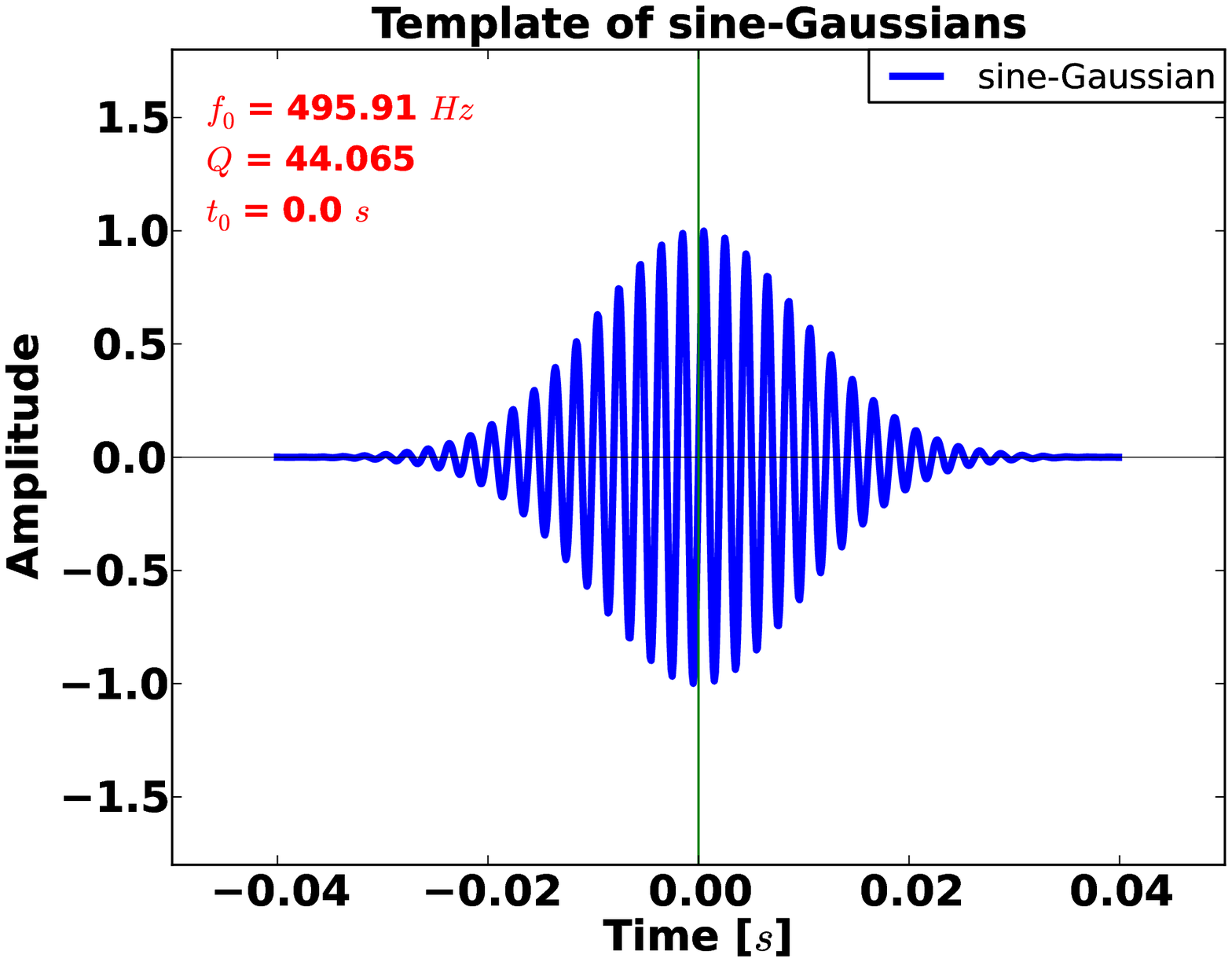}}&
\subfloat[Part 2][A sample Chirplet.]{\includegraphics[width=.45\textwidth,height=0.25\textheight]{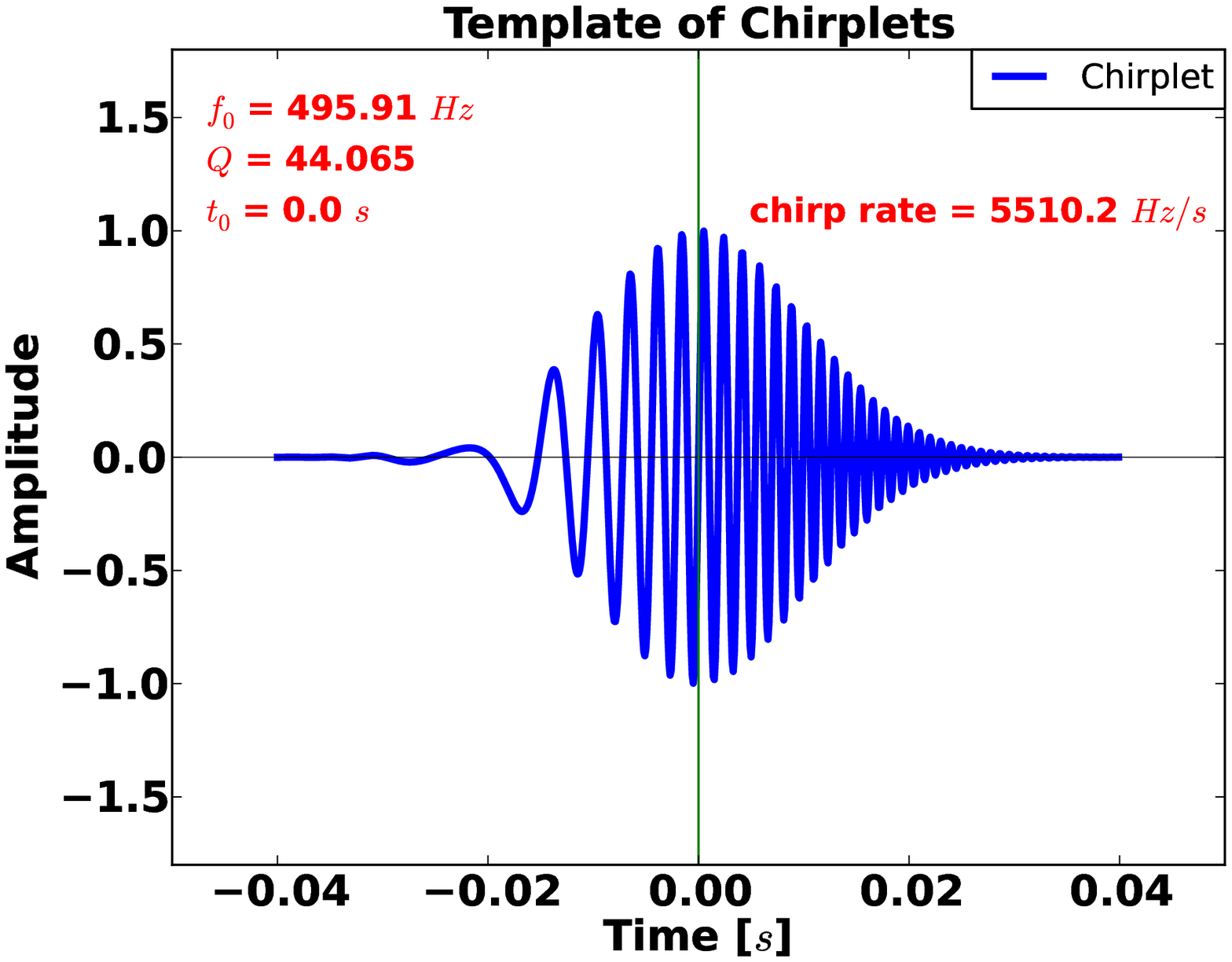}}
\end{tabular}
\caption{\label{sg_chirplet}\textbf{Sine-Gaussian and Chirplet waveform:} (\textit{left}) Example of a Sine-Gaussian waveform. (\textit{right}) Example of a Chirplet waveform.}
\end{figure}

The sine-Gaussian template bank used in the standard Omega burst search, are defined in the time domain as:
\begin{equation}
\psi(\tau)\equiv A \exp \left( -\frac{(2\pi f)^2}{Q^2} (\tau-t)^2\right)
\exp \left( 2\pi i \left[ f (\tau-t) \right]\right),
\end{equation}
Chirplets are a generalization of  sine-Gaussians, defined in time domain as:
\begin{equation}
\psi(\tau)\equiv A \exp \left( -\frac{(2\pi f)^2}{Q^2} (\tau-t)^2\right)
\exp \left( 2\pi i \left[ f (\tau-t) + d/2\: (\tau-t)^2 \right]\right),
\end{equation}
where $A=(8\pi f^2/Q^2)^{1/4}$ is a normalization factor ensuring that $\int
|\psi|^2 d\tau=1$; $t$ and $f$ are the central time and central frequency, respectively and $Q$
is the dimensionless quality factor. See Fig. \ref{sg_chirplet} for an example of sine-Gaussian and a chirplet.

The difference between a chirplet and a  sine-Gaussian waveform is the \textit{chirp rate}, an additional term in the phase, denoted as $d$, that linearly controls the slope of the chirplet frequency evolution as $f(\tau)=f+d (\tau-t)$.  When $d=0$, a chirplet becomes the standard sine-Gaussians. Chirplets are thus described with a four-dimensional parameter space of $\{t, f, Q, d\}$.

The \textit{chirplet transform} $T$ is obtained by correlating the data with chirplets. In the frequency domain, it reads:
\begin{equation}
  \label{ct}
  T[x;\{t, f, Q, d\}]= \left| \int X(\xi) \Psi^*(\xi;\{t, f, Q, d\}) d\xi \right|^2,
\end{equation}
where $X(\xi)$ and $\Psi(\xi;\{t, f, Q, d\})$ denote the Fourier transforms of
the (whitened) data stream $x(\tau)$ and chirplet $\psi(\tau)$ with the parameters $\{t, f, Q, d\}$ respectively.

\section{Performances of Chirpletized Omega}

\subsection{Searching for binary coalescence with chirplets}

The gravitational wave spectrum from the coalescence of a binary system is at lower frequency for system with larger total mass. In particular, the frequency associated with the inner-most stable circular orbit (ISCO), which corresponds to the transition between inspiral and merger phase of the coalescence, is below 80 Hz for systems with total mass $M \gtrsim 100 M_{\odot}$.  With this feature, the chirp phase of the waveform with a spectral content at frequencies below ISCO is outside the detector sensitive band and thus does not contribute significantly to the total detected SNR of the signal. Sine-Gaussian waveforms are a reasonable approximation for the few waveform cycles associated with the merger and ringdown portions of the coalescence waveform. 
For a symmetric system with lower mass ($M < 30 M_{\odot}$), the coalescence time of BBH in a ground based detector's sensitive band is $>$ 400 ms. There is loss of measured signal-to-noise ratio induced by the mismatch in the waveforms; between the chirp nature of the GW evolution and the constant central frequency sine-Gaussian template bank. The linear frequency variation of chirplets provide better match to the chirping signal of BBH in this mass range.

\subsection{Increased signal-to-noise ratio}

\begin{figure}
\begin{tabular}{cc}

\subfloat[Part 1][Eventgram in sine-Gaussian bank.]{\includegraphics[width=.45\textwidth,height=0.2\textheight]{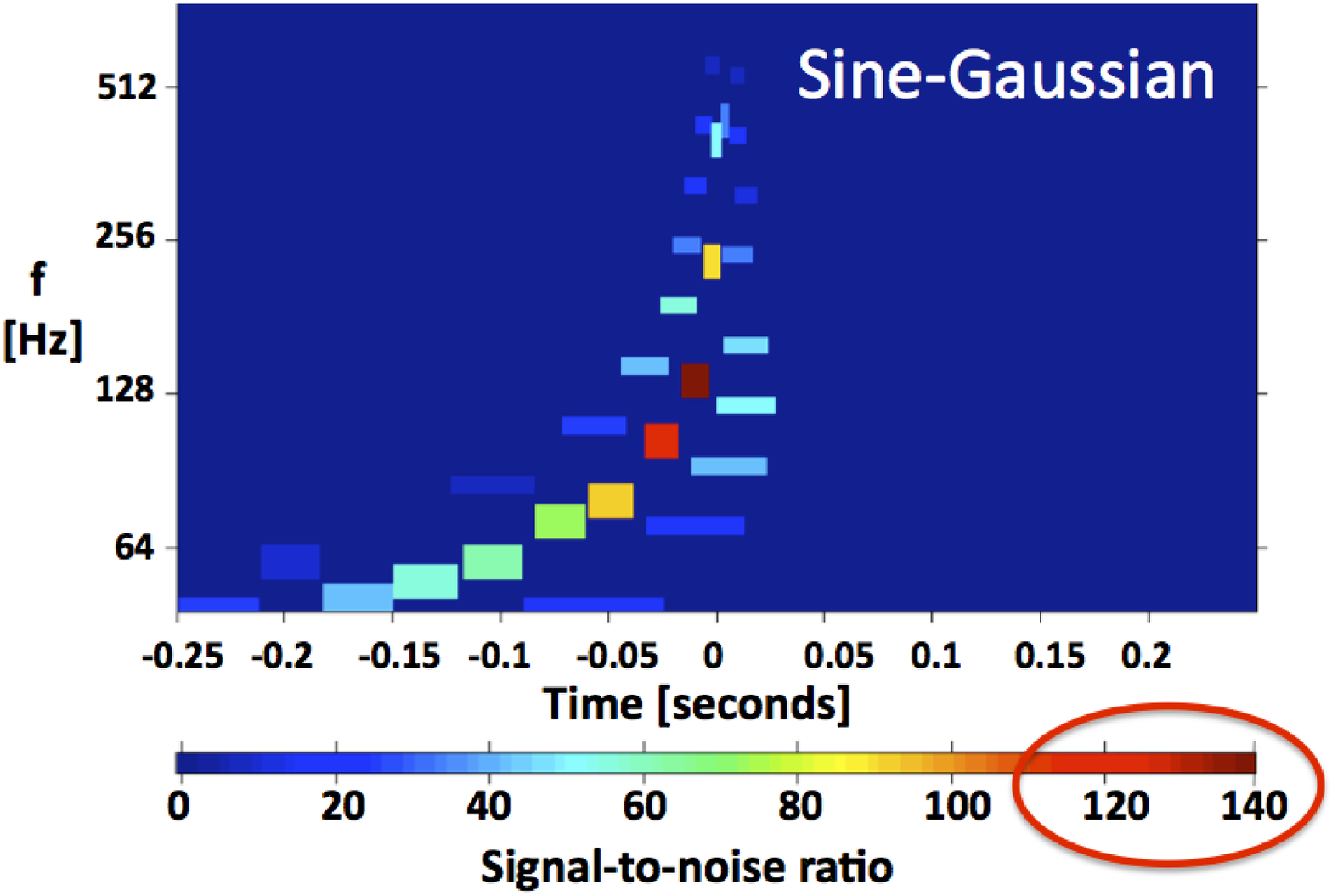}}&
\subfloat[Part 2][Eventgram in chirplet bank.]{\includegraphics[width=.45\textwidth,height=0.2\textheight]{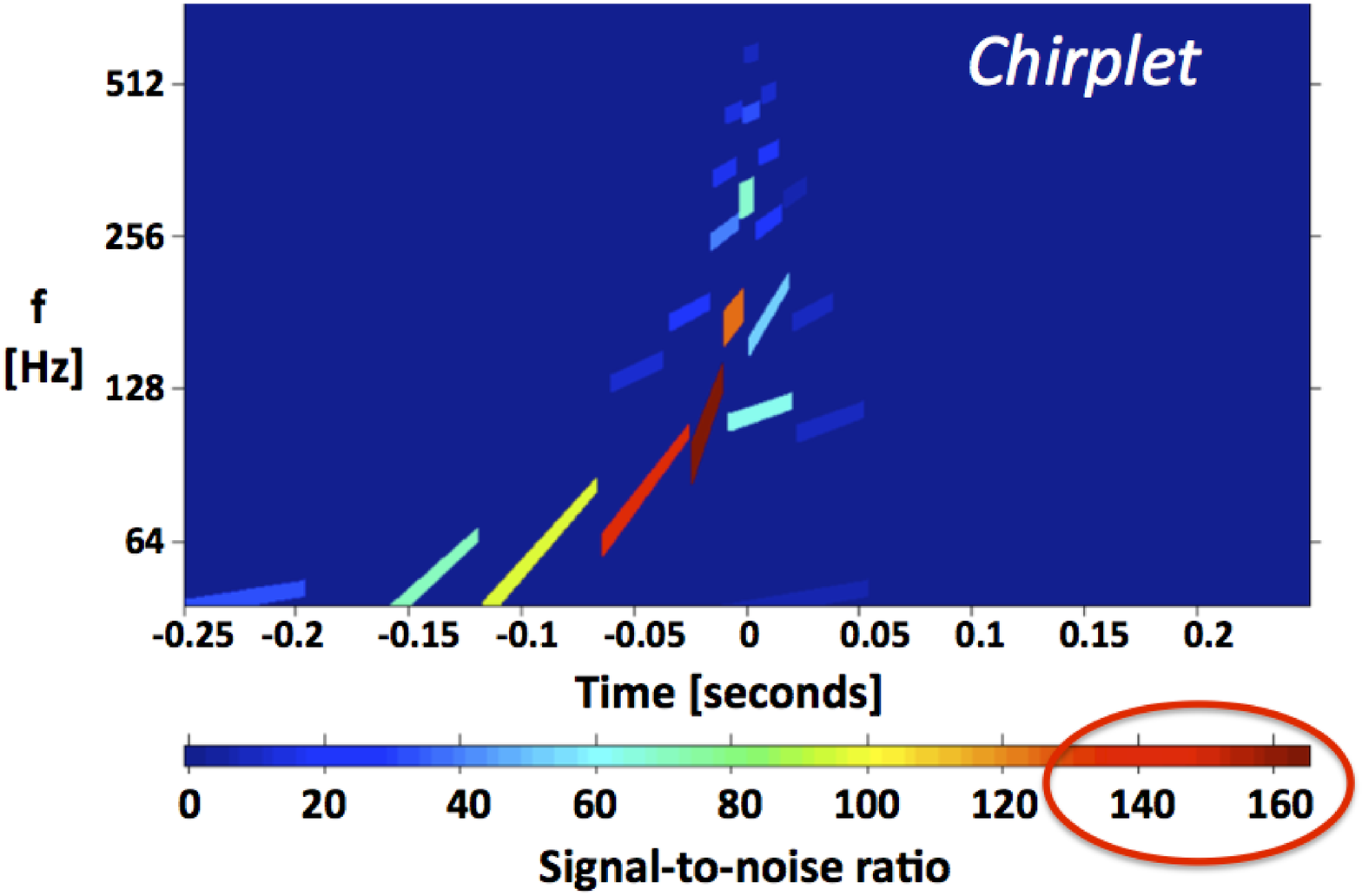}}
\end{tabular}
\caption{\label{eventgram}\textbf{Eventgrams for sine-Gaussian and Chirplet banks:} The eventgrams are plotted for inspiralling black-hole binary (with $m_1=13 M_{\odot}$ and $m_2=20 M_{\odot}$) signal in simulated Gaussian LIGO noise. See the difference between the constant frequency tiles for the sine-Gaussians and the linear frequency variation for chirplets. An increase of SNR by $\sim$ 20\% in the peak pixel (the most significant time-frequency pixel) is to be noted.}
\end{figure}

It has been shown~\cite{chirplet-paper} that Chirpletized Omega measures a higher matched filtering signal-to-noise-ratio (SNR) for BBH signal $< 100 M_{\odot}$. The SNR improvement is ($\sim 20 \%$) for the mass range $M \lesssim 60 M_{\odot}$ and the improvement may go up-to $\sim 40\%$ for  $M \lesssim 20 M_{\odot}$. When compared to actual matched filtering SNR of signals, chirplet recovers the full SNR for $M \lesssim 20 M_{\odot}$.

As an illustration, we show an eventgram (Fig.~\ref{eventgram}). An eventgram shows the significant time-frequency pixels for different $Q$ values. In this particular eventgram a BBH signal is added to Gaussian noise with the  spectral characteristics of initial LIGO design.  The signal is a non-spinning BBH system with mass components  $m_1=13 M_{\odot}$ and $m_2=20 M_{\odot}$. In this example, chirplets with a positive slope are chosen to sine-Gaussian with constant frequency and the correlation of the most significative chirplet is,  $\sim 20 \%$  larger than the most significative sine-Gaussian.

\subsection{Increased detectability}

The detectability of BBH signals is limited by a background of noise fluctuations that trigger the search algorithms. Noise fluctuations can be reduced by eliminating non-GW triggers from the search  with the application of a suitable coincident or coherent strategy for a network of GW detectors. The background in a chirplet search is in general different from the typical background in the standard Omega burst search. An additional 5\% increase was reported~\cite{chirplet-paper} in the measured single detector SNR for background triggers in a chirplet burst search compared to standard Omega. Previous analysis~\cite{chirplet-paper} was conducted with different false alarm rates. In order to ensure an unbiased comparison, this paper addresses this issue and compares the detectability in the two search algorithms at the same false alarm rate (FAR). In the current study we added BBH signals to 2 months of simulated colored Gaussian noise mimicking the instrumental noise of the two Hanford LIGO detectors (H1 and H2), and the Livingston LIGO detector (L1). We split the range of BBH masses in two sets, total mass 4-35 $M_{\odot}$ and total mass 35-80 $M_{\odot}$. The distribution of mass was uniform in the component masses of the two black holes. We estimated the network's background  by time-sliding the triggers from L1 detector with the coincident triggers from H1 and H2. Similarly foreground triggers were collected by coincidence of triggers from H1, H2 and L1. A time window of 10 ms was used to find the coincident triggers. The times at which the signals were added to noise are known. We used those times to identify significant events within 10 ms in the foreground triggers.
We ranked the triggers from both the background and the simulated signals using the the geometric mean of the single detector signal-to-noise ratios as ranking statistics. 

\begin{equation}
\rho_{GM} = \left(\rho_{H1} \times \rho_{H2} \times \rho_{L1}\right)^{\frac{1}{3}} ,
\end{equation}
where $\rho_{detector}$ is the single detector SNR.

After all the triggers are ranked, the significance of each trigger was evaluated by computing its probability to belong to background (FAR). The performance comparison between chirplet and standard Omega searches was done with a common threshold on the FAR.

\begin{figure}
\begin{tabular}{cc}
\subfloat[Part 1][Efficiency curves for set 4-35 $M_\odot$.]{\includegraphics[width=.47\textwidth]{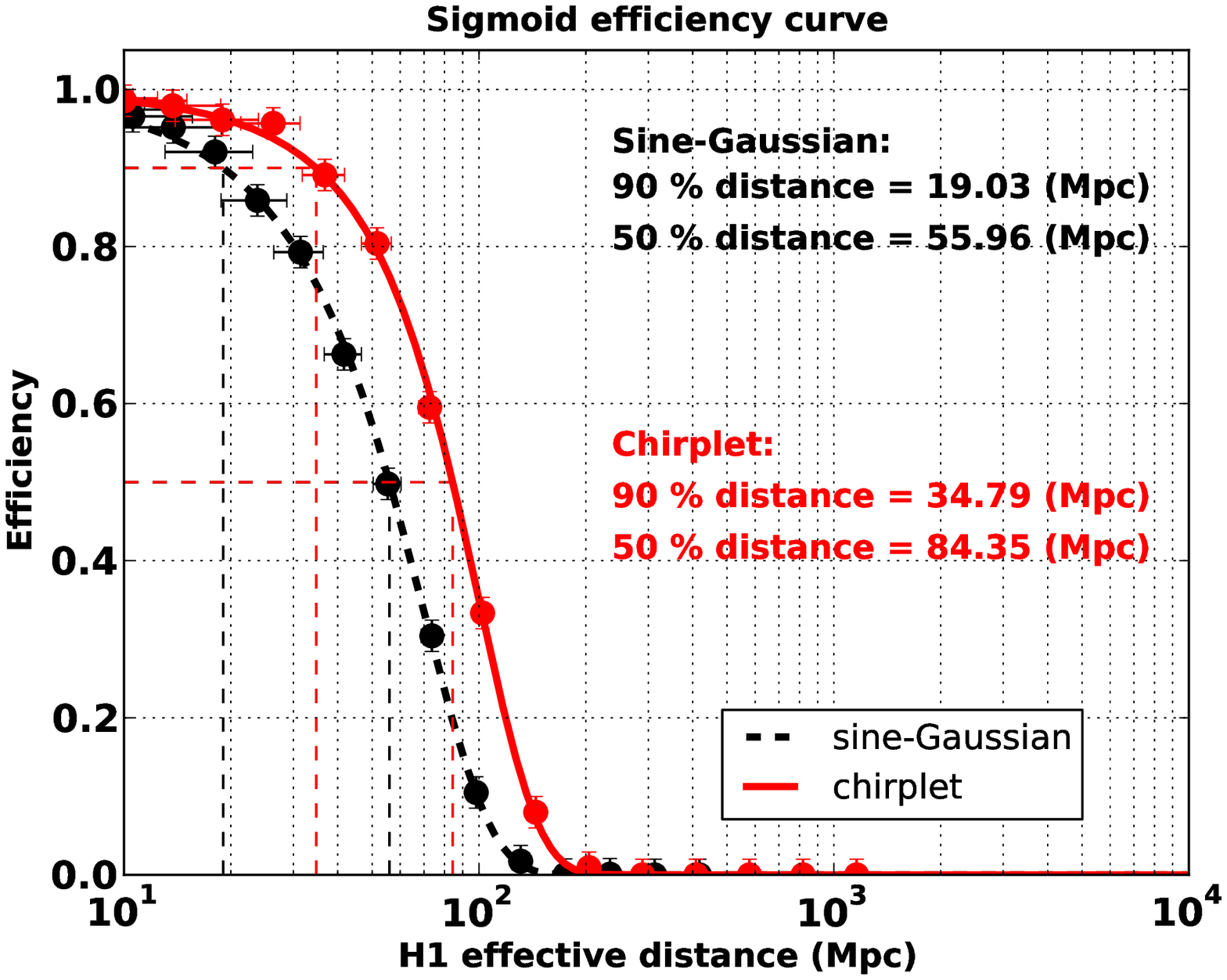}\label{fig:omegaSETA}}&
\subfloat[Part 2][Efficiency curves for set 35-80 $M_\odot$.]{\includegraphics[width=.47\textwidth]{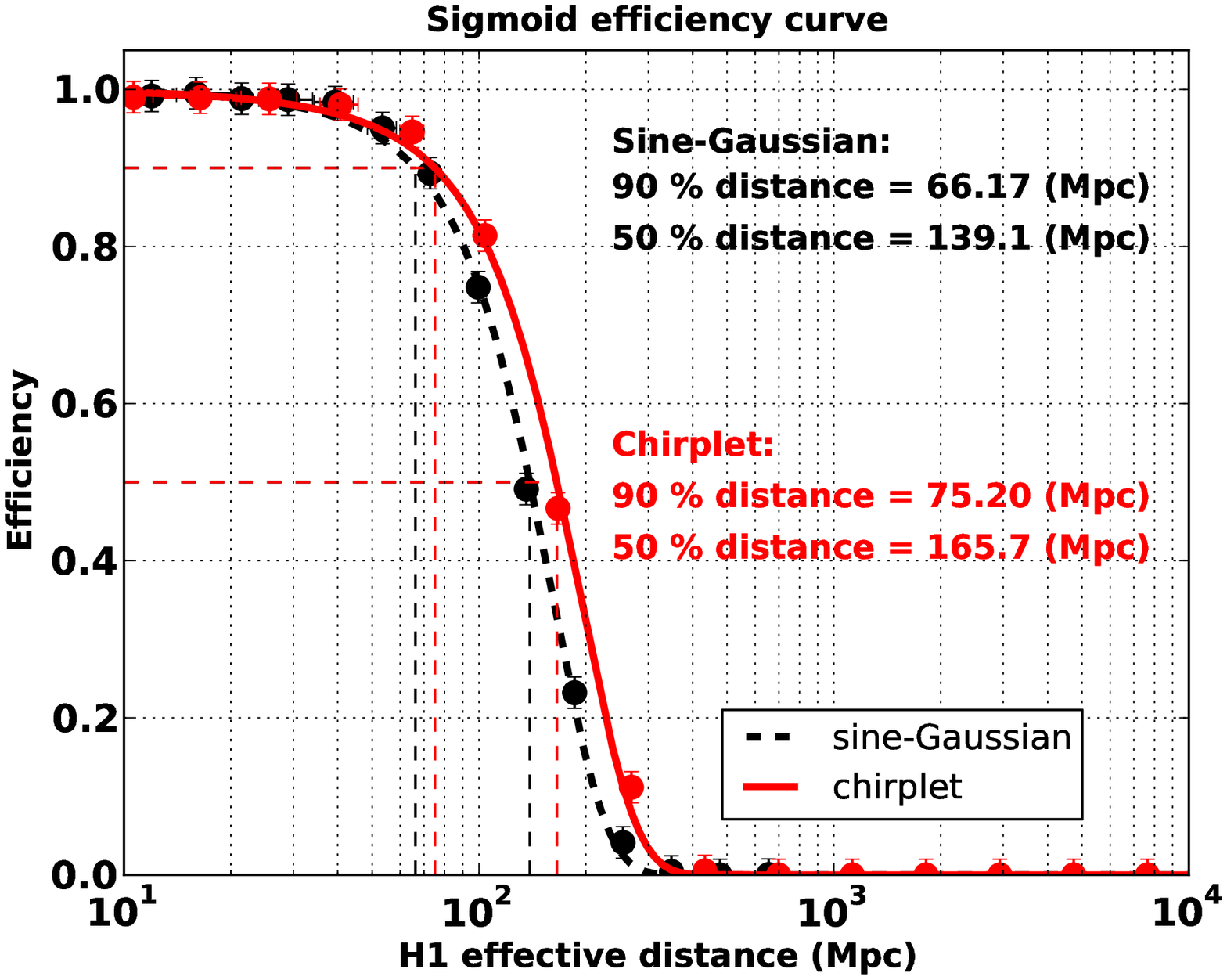}\label{fig:chirpletSETA}}
\end{tabular}
\caption{\label{sigmoid}\textbf{Efficiency curve:} Efficiency sigmoids are plotted with a FAR of $10^{-8}$ Hz. Higher detection efficiency in a chirplet search is observed. The improvement in chirplet is more pronounced for the set with mass 4-35 $M_\odot$.}
\end{figure}

\newpage
\paragraph{Efficiency curve ---}
We compared the performance of the sine-Gaussian and the Chirplet algorithms by estimating the detection efficiency as a function of the effective distance of the BBH from the detector. We fit a sigmoid to the observed number of signals as a function of the distance, at a fixed false alarm rate of $10^{-8}$ Hz. The characteristic distance parameters, known as the 50\% and 90\% efficiency distances, which are the points on the sigmoid where 50\% and 90\% of the signals are recovered respectively, are both improved for chirplet compared to Omega as seen in Fig.~\ref{sigmoid}. The numbers obtained from the sigmoid fit are shown below in Table 1. The overall improvement of $>$ 50\% in the distance parameters, is higher for the lower mass set (4-35 $M_{\odot}$) compared to, up-to 40\% for the mass set with total mass ranging 35-80 $M_{\odot}$.

\begin{table}[h]
\begin{center}
\begin{tabular}{ | c | c | c | c | c | }
  \hline
   & Sine-Gaussians & Sine-Gaussians & Chirplets & Chirplets\\
  Set & 50\%  (Mpc) & 90\%  (Mpc) & 50\%  (Mpc) & 90\%  (Mpc)\\
  \hline
  4-35 $M_{\odot}$ & 56 $\pm$ 4 & 19 $\pm$ 1 & 84 $\pm$ 2 & 35 $\pm$ 2 \\
  \hline
  35-80 $M_{\odot}$ & 139 $\pm$ 4 & 66 $\pm$ 2 & 166 $\pm$ 4 & 75 $\pm$ 3 \\
  \hline
\end{tabular}
\end{center}
\caption{50\% and 90\% efficiency distance at a FAR of $10^{-8}$ Hz.}
\end{table}
\normalsize

\begin{figure}
\begin{center}
\hspace{-1em}
\subfloat[Part 1][Sine-Gaussians on 4-35 $M_\odot$.]{\includegraphics[width=.34\textwidth,height=0.34\textwidth]{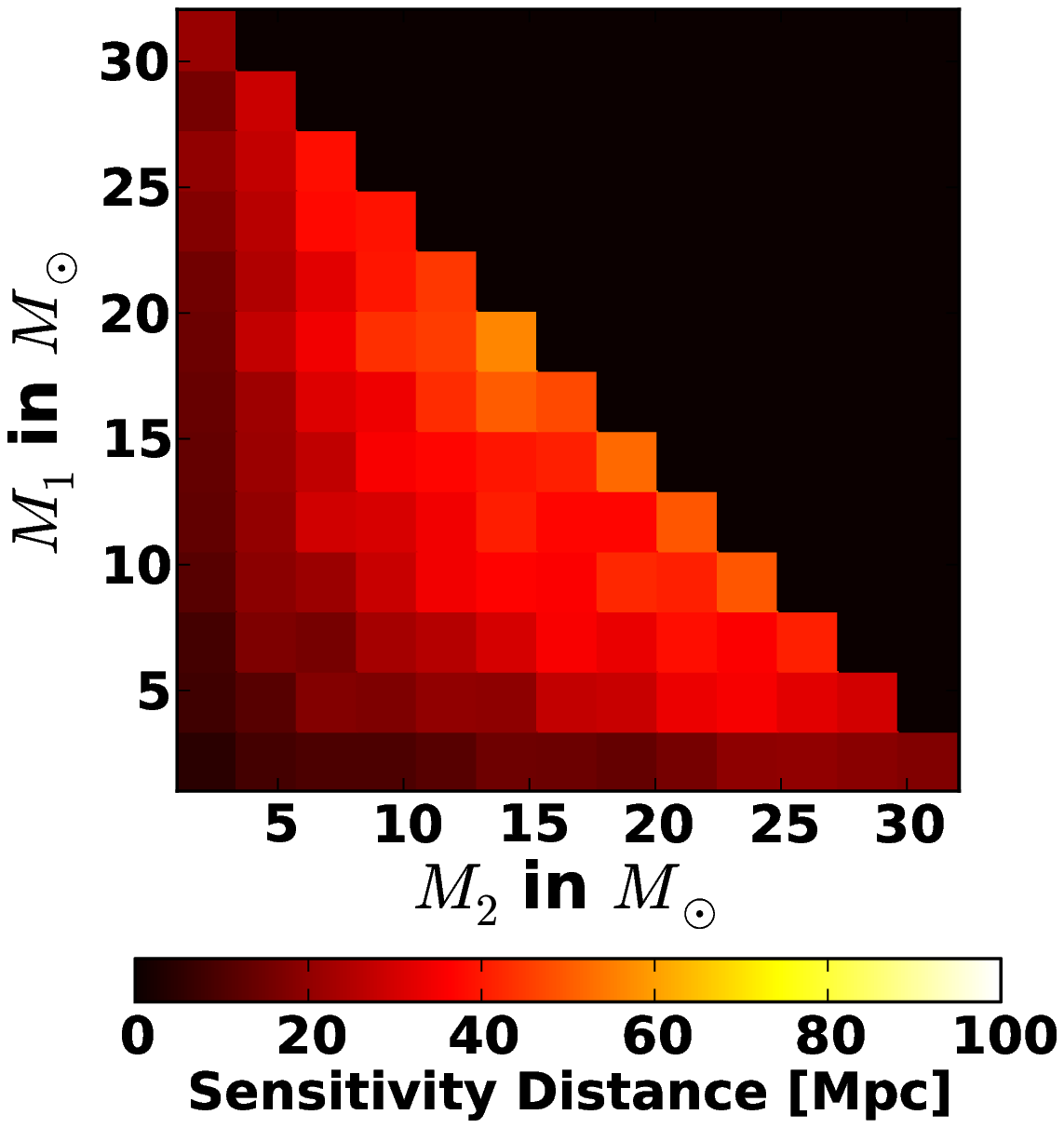}\label{fig:omegaSETA}}
\subfloat[Part 2][Chirplets on 4-35 $M_\odot$.]{\includegraphics[width=.34\textwidth,height=0.34\textwidth]{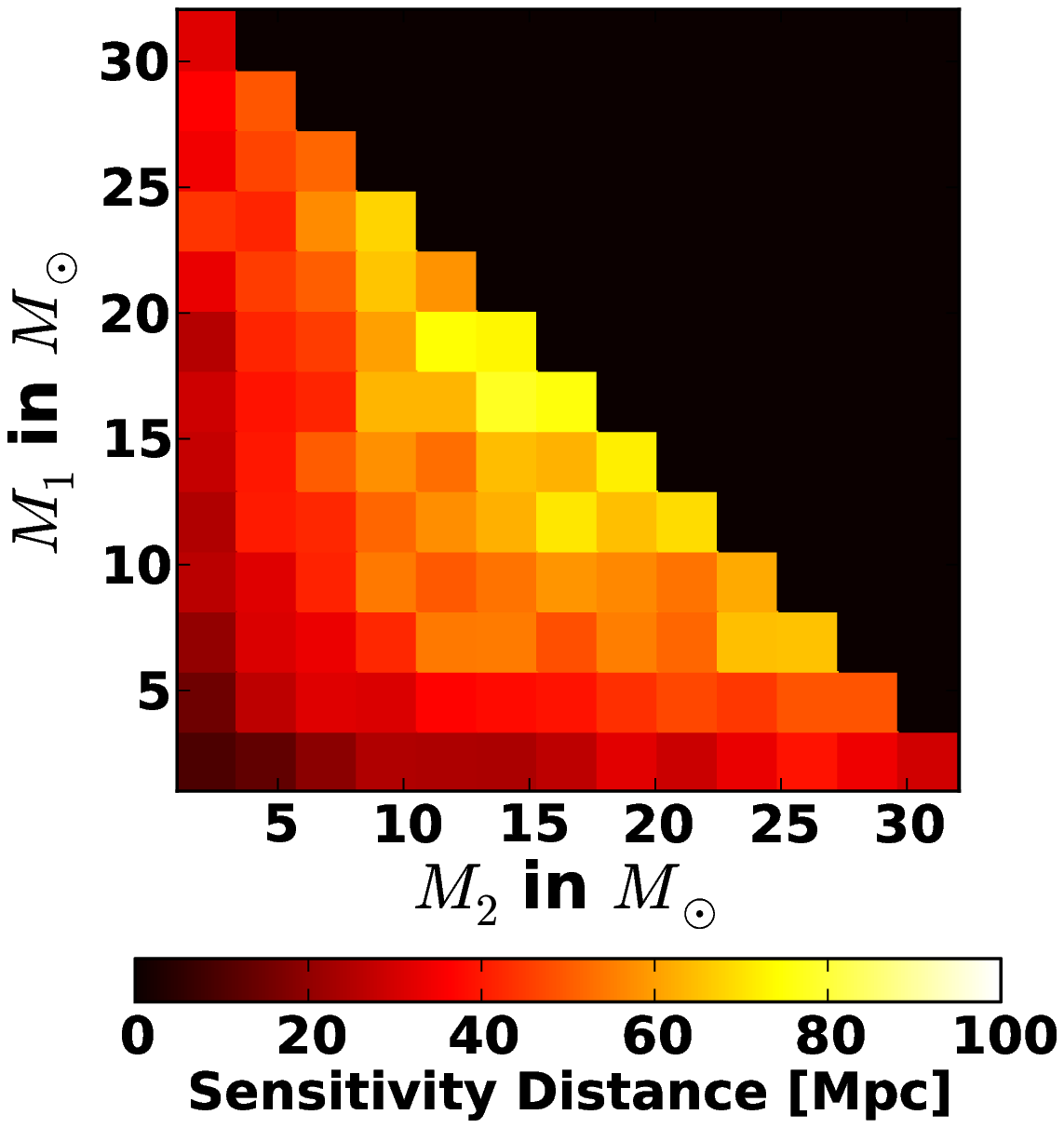}\label{fig:chirpletSETA}}
\subfloat[Part 3][Chirplet/sine-Gaussian:4-35 $M_\odot$.]{\includegraphics[width=.34\textwidth,height=0.34\textwidth]{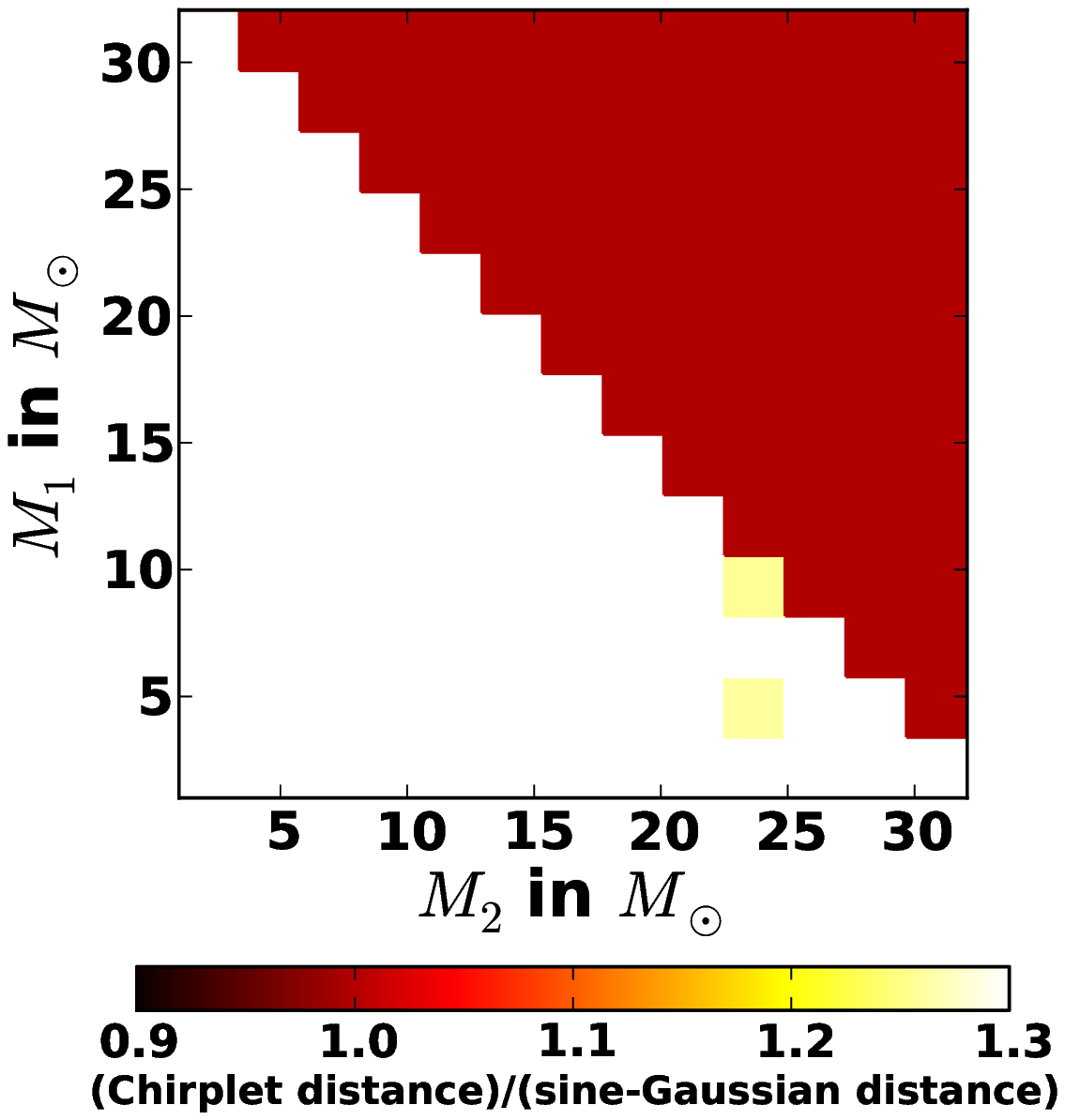}\label{fig:ratioSETA}}\\
\hspace{-1em}
\subfloat[Part 4][Sine-Gaussians on 35-80 $M_\odot$.]{\includegraphics[width=.34\textwidth,height=0.34\textwidth]{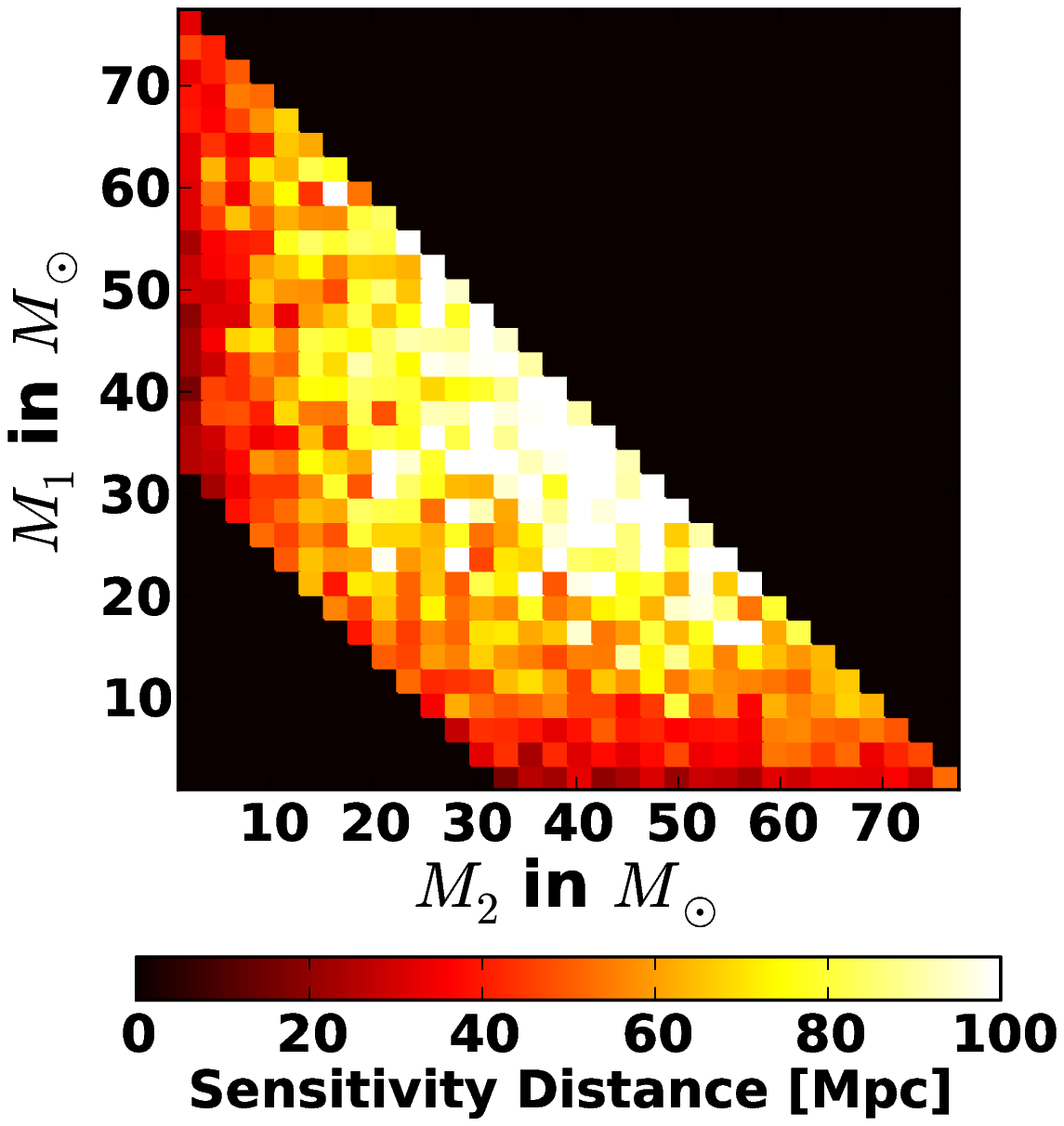}\label{fig:omegaSETB}}
\subfloat[Part 5][Chirplets on 35-80 $M_\odot$.]{\includegraphics[width=.34\textwidth,height=0.34\textwidth]{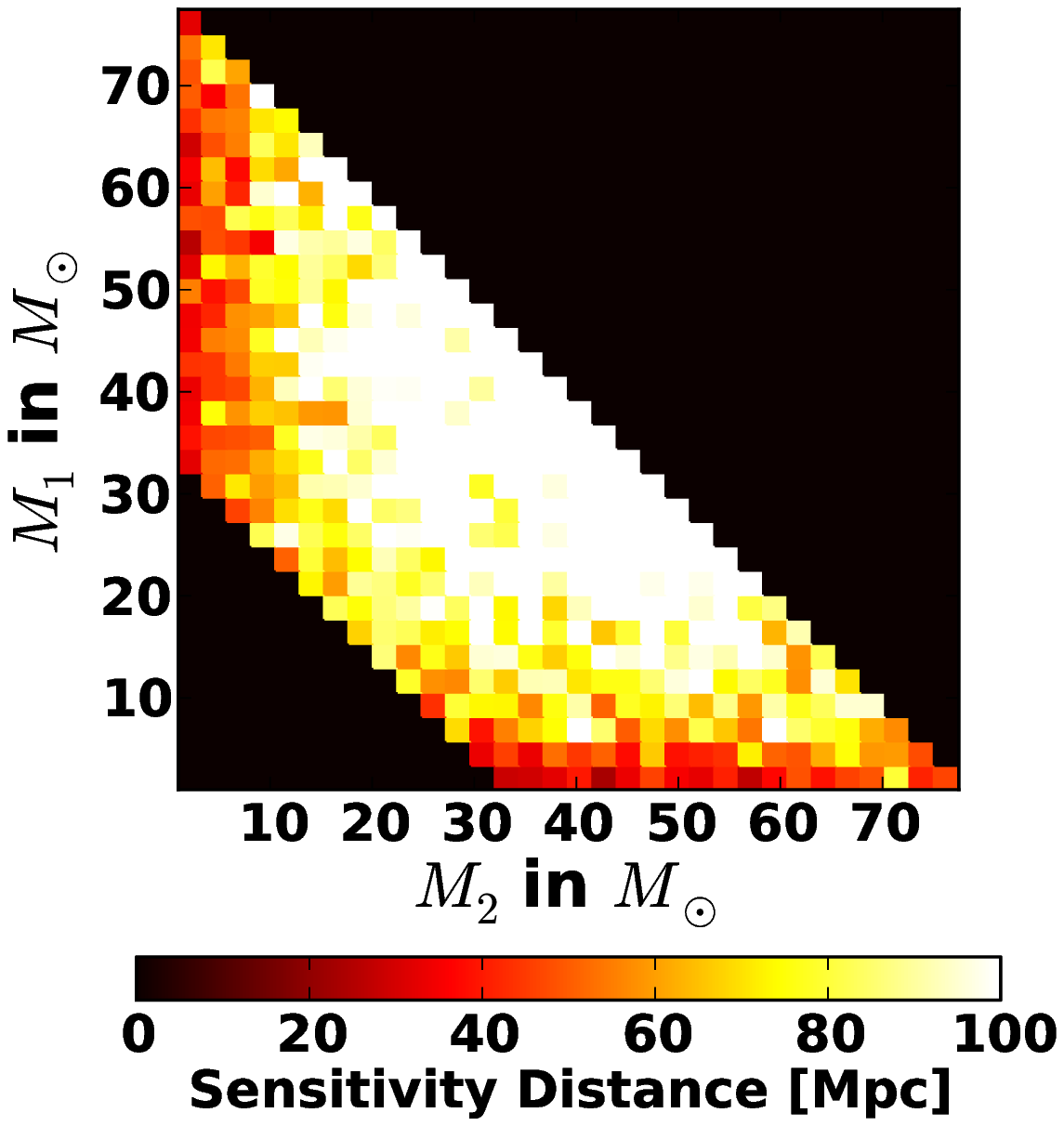}\label{fig:chirpletSETB}}
\subfloat[Part 6][Chirplet/sine-Gaussian:35-80 $M_\odot$.]{\includegraphics[width=.34\textwidth,height=0.34\textwidth]{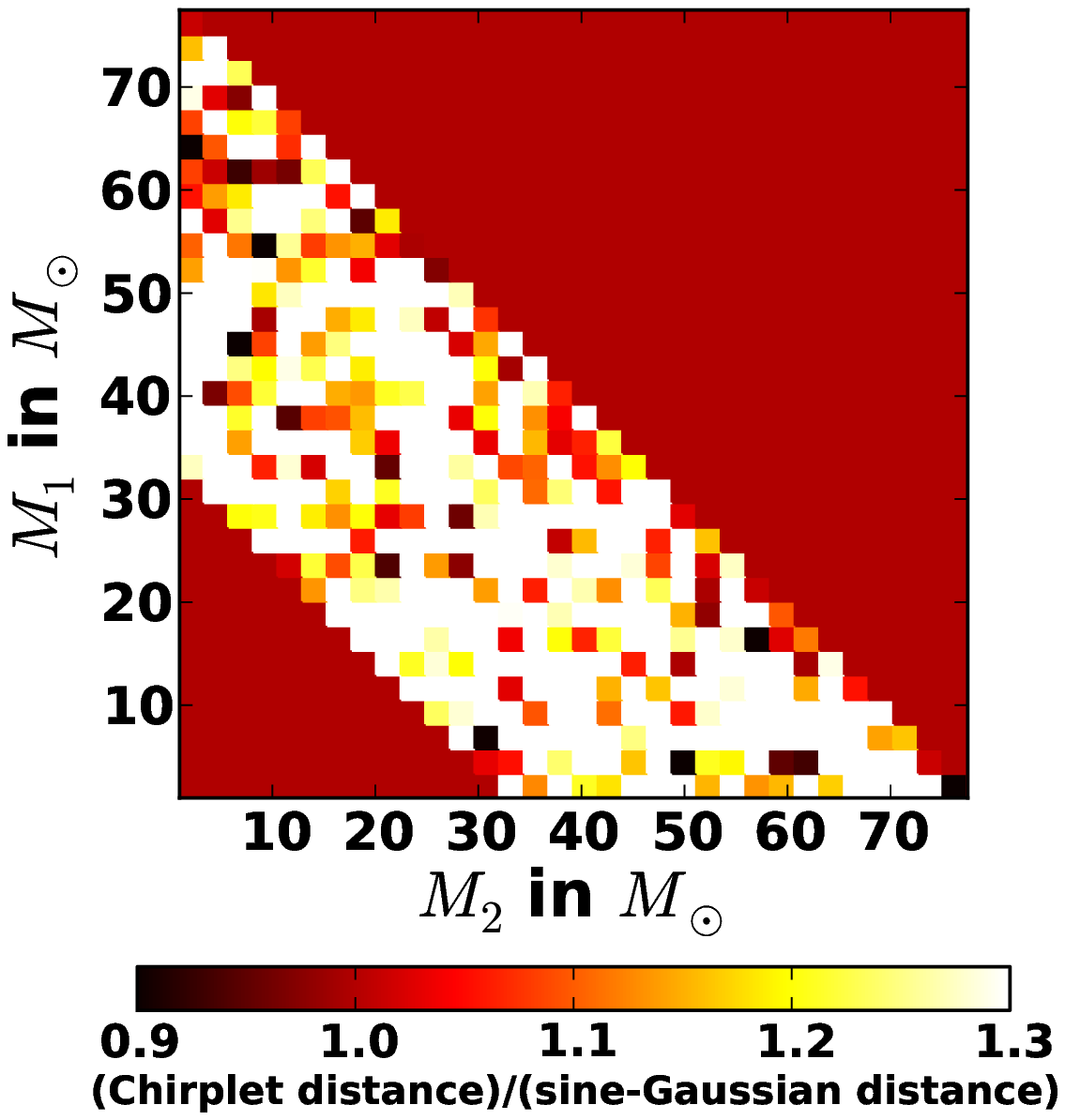}\label{fig:ratioSETB}}
\end{center}
\caption{\label{sensitivity}\textbf{Sensitivity distance:} Plots of the fiducial distance that can be observed at a FAR of $10^{-8}$ Hz for both chirplet and sine-Gaussian in the mass parameter space are shown in Fig. ~\ref{fig:omegaSETA},~\ref{fig:chirpletSETA},~\ref{fig:omegaSETB} and ~\ref{fig:chirpletSETB}. The ratio plots of the detectable distance are shown in Fig.~\ref{fig:ratioSETA} and ~\ref{fig:ratioSETB}. \textit{More than} 30\% of improvement in the sensitivity distance for the set 4-35 $M_\odot$ in all the mass bins and \textit{up-to} 30\% improvement in the sensitivity distance for most of the mass bins in set 35-80 $M_\odot$ should be noted.}
\end{figure}

\paragraph{Sensitivity distance ---}
While efficiency sigmoids are good measures for the detection performance, as shown in Fig.~\ref{sigmoid}, they don't provide details of how detectability changes across mass bins. For this reason, we compared the fiducial distance the detector network is sensitive at a false alarm rate of $10^{-8}$ Hz, or the {\it sensitivity distance}, as the function of the BBH mass parameters. Fig.~\ref{fig:omegaSETA} and  Fig.~\ref{fig:chirpletSETA} show sensitivity distance for chirplet and standard Omega search in the mass parameter space for the set with total mass 4-35 $M_\odot$. Fig.~\ref{fig:ratioSETA} shows the ratio of the sensitivity distance measured by chirplet to the sensitivity distance measured by standard Omega search. For this set, chirplet search improves detectability \textit{more than} 30\%  over standard Omega search in all mass bins as seen in the ratio plot of Fig.~\ref{fig:ratioSETA}. Similarly, Fig.~\ref{fig:omegaSETB} and  Fig.~\ref{fig:chirpletSETB} show detectability for set with total mass 35-80 $M_\odot$. In this set the detectability in chirplet search also improves \textit{up-to} 30\% over standard Omega search for most of the mass bins and is shown in Fig.~\ref{fig:ratioSETB}. 

\paragraph{}
The above comparisons of efficiency sigmoids and sensitivity distances at a false alarm rate of $10^{-8}$ Hz point very well to higher detectability of BBH signal in a chirplet burst search over standard Omega burst search, specially in the lower mass region (total mass $<$ 35 $M_{\odot}$). This is in accordance with the previous result of increased SNR~\cite{chirplet-paper}.

\section{Future plans}

In this study, we have demonstrated the advantage of a chirplet burst search over the standard sine-Gaussian burst search using coincidence method as the multi-detector network strategy. There is room for additional modifications in the chirplet-based analysis, with the implementation of a coherent detection statistics and novel clustering strategies, which are currently under study. There have been studies~\cite{svd1,svd2} based on matrix factorization techniques to construct orthonormal bases that span the space of high-mass merger waveforms.  Using chirplets as templates, is a similar but distinct approach, with different advantages.  In particular, a chirplet based search is not tied to the availability of an explicit representation of the family of waveforms to construct the basis, which gives it greater simplicity and robustness in the face of possible errors in the numerical relativity simulations, although, in exchange, it suffers some possible SNR loss.

In addition to an improved detectability, the chirplet-based analysis offers additional information on the frequency evolution of the observed signal, which will be useful in the \textit{a posteriori} interpretation of an event.

The work presented in this paper does not address the question of the spin of the binary black hole system. Spins affect the detectability profoundly~\cite{spin-phenom}. We are planning to extend our current studies to express the detectability measured in a chirplet burst search to spin parameter space.

\section*{References}

\bibliographystyle{unsrt}
\bibliography{chirplet}

\ack{}
We are grateful to the LIGO Scientific Collaboration for the use of its
algorithm library available at
\url{https://www.lsc-group.phys.uwm.edu/daswg/projects/lalapps.html}.  This work
is supported in part by NSF grant PHY-0653550.  
\begin{center}
\vspace{0.5em}
This document has been assigned LIGO Laboratory document number P1100138.
\end{center}
\end{document}